\documentclass[epj]{svjour}
\usepackage{epsfig}
\usepackage{color}
\usepackage{amsmath}

\def\Id{{\rm 1\kern-.3em I}}

  \newbox\charbox
  \newbox\slabox
  \def\s#1{{      % Feynman slash
        \setbox\charbox=\hbox{$#1$}
        \setbox\slabox=\hbox{$/$}
        \dimen\charbox=\ht\slabox
        \advance\dimen\charbox by -\dp\slabox
        \advance\dimen\charbox by -\ht\charbox
        \advance\dimen\charbox by \dp\charbox
        \divide\dimen\charbox by 2
        \raise-\dimen\charbox\hbox to \wd\charbox{\hss/\hss}
        \llap{$#1$}
        }}
\begin{document}

\title{Semileptonic decays of baryons in a relativistic quark model}
\author{Sascha Migura \and Dirk Merten \and Bernard Metsch \and Herbert-R.~Petry}
%\author{First author\inst{1} \and Second author\inst{2}% etc
\institute{Helmholtz-Institut f\"ur Strahlen- und Kernphysik, Nu{\ss}allee 14--16, 53115 Bonn, Germany \\ \email{migura@itkp.uni-bonn.de}}
%\institute{Insert the first address here \and the second here}
%
\date{Received: 16 February 2006 /}
\abstract{
We calculate semileptonic decays of light and heavy baryons in a relativistically covariant constituent quark model. The model is based on the Bethe-Salpeter-equation in instantaneous approximation. It generates satisfactory mass spectra for mesons and baryons up to the highest observable energies. Without introducing additional free parameters we compute on this basis helicity amplitudes of electronic and muonic semileptonic decays of baryons. We thus obtain form factor ratios and decay rates in good agreement with experiment.
 % end of PACS codes
} %end of abstract
\maketitle

\section{Introduction}
The concept of constituent quarks is so far the most successful approach when describing hadronic spectra. It is assumed that constituent quarks form due to spontaneous breaking of chiral symmetry of QCD but this assumption is so far difficult to be verified quantitatively in a consistent way. What can be done, however, is to investigate up to which resonance energies this concept is valid in comparison to experiment and at which masses it fails.

This paper is part of a series which systematically study the whole hadron resonance region within a unified constituent quark model with a minimal number of free parameters. Based on the Bethe-Salpeter-equation the model is relativistically covariant by construction. It allows correct boosting prescriptions and is especially important in the present work where we investigate e.g.~a heavy baryon decaying into a light baryon and where large recoils hence play a decisive role.

The speciality of our Bethe-Salpeter-model is the inclusion of 't Hooft's instanton-induced force, see \cite{'tHooft:1976fv,Shifman:1979uw} for details. In combination with a linearly rising confinement potential 't Hooft's force leads to much better spectroscopic results than the conventional one-gluon-exchange within the same framework.

In particular, the Bethe-Salpeter-model is able to describe the complete spectra of light mesons offering a natural solution to the $U_A(1)$-problem \cite{Koll:2000ke,Ricken:2000kf}. But also meson decay properties like form factors, decay constants and decay widths for two-photon-decays and electromagnetic form factors have been investigated \cite{Koll:2000ke}. Form factors for semileptonic decays of heavy mesons have been evaluated in very good agreement with experimental data \cite{Merten:2001er}. Even strong two-body decays of light mesons have been studied \cite{Ricken:2003ua}.  

The Bethe-Salpeter-model has also been successfully applied to the calculation of baryon properties. A detailed analysis of the complete light baryon spectra has been presented \cite{Loring:2001kv,Loring:2001kx,Loring:2001ky}. The Regge-trajectories up to the highest observed energies and the hyperfine structures are correctly reproduced. On the basis of these results electroweak form factors and photon induced transitions of non-strange and strange baryons have been calculated in good agreement with experiment \cite{Merten:2002nz,VanCauteren:2005sm}.

The present paper adds to these achievements a calculation of form factor ratios and decay rates of semileptonic decays of baryons. As before this is done without introducing additional free parameters which are completely fixed by the baryon spectra alone. The results are therefore absolute predictions.

We consider semileptonic decays of the octet- and the decuplet-baryons. These are decays of light baryons outside the scope of heavy quark effective theories. For comparison we also compute the semileptonic decay of a heavy, charmed baryon, namely the $\Lambda_c$. We calculate in each case both, electronic and muonic decays. 

These are all crucial calculations for testing our model also at large hadron masses and hence the validity of the constituent quark picture since in a unified description of baryons we are not only interested in quantities describing a single baryon but the complete hadron spectrum including transitions. This is in particular interesting with regard to the interplay between weak and strong interactions.

The paper is organized as follows: Section~\ref{sec:BSM} recapitulates briefly the Bethe-Salpeter-model. It explains the ingredients and basic equations of the model and shows how current-induced matrix elements can be calculated. Section~\ref{sec:BasicsSemilep} briefly reviews the underlying theory of semileptonic decays. We restrict ourselves to those parts of the electroweak theory needed to derive formulas for the physical quantities which we then evaluate numerically. Section \ref{sec:results} shows our theoretical values for form factor ratios, differential and total decay rates. As far as possible these theoretical values are compared to experimental data. Otherwise they are stated as predictions. Section~\ref{sec:conclusion} concludes.

\section{Bethe-Salpeter-model} \label{sec:BSM} 
\subsection{Bound states}
Bound states are essentially a non-perturbative phe\-no\-me\-non and the Bethe-Salpeter-equation offers a suitable starting point. The usual quantum mechanical wave function used by many other quark models is replaced by the Bethe-Salpeter-amplitude $\chi$ that describes the baryonic bound states. It is defined via quark field operators $\Psi_{a_i}(x_i)$ by
\begin{eqnarray}
\chi_{\bar P\,a^{}_1 a^{}_2 a^{}_3}(x^{}_1,x^{}_2,x^{}_3)
=\langle0|T\Psi^{}_{a^{}_{1}}(x^{}_{1})\Psi^{}_{a^{}_{2}}(x^{}_{2})
\Psi^{}_{a^{}_{3}}(x^{}_{3})|\bar P\rangle \label{eqn:BSA}
\end{eqnarray}
where $\bar P$ denotes the four-momentum of the on-shell bound state and $T$ the time ordering operator. Due to translational invariance it is advantageous to switch from now on to relative coordinates $p_\xi$ and $p_\eta$ in momentum space.

\begin{figure}[t]
\begin{center}
\input{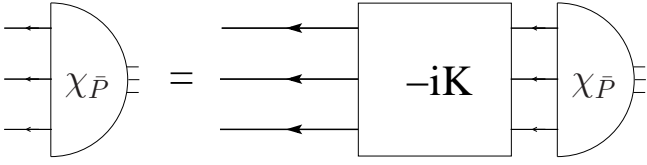}
\caption{Graphical illustration of the three-quark Bethe-Salpeter-equation (\ref{eqn:BSE}). Thick arrows indicate the full quark propagators $S^i_F$. Irreducible parts $K$ are absorbed into the Bethe-Salpeter-amplitude $\chi_{\bar P}$.}
\label{fig:BSE}
\end{center}
\end{figure}

The Fourier-transform of the Bethe-Salpeter-amplitude is determined by the Bethe-Salpeter-equation
\begin{eqnarray}
&& \chi_{\bar P\,a_1 a_2 a_3}(p_\xi,p_\eta)=
S^1_{F\,a_1 a'_1}\left(\textstyle \frac{1}{3}\bar P+p_{\xi}+\frac{1}{2}p_{\eta}\right)\, \nonumber \\
&& \,\times S^2_{F\,a_2 a'_2}\left(\textstyle \frac{1}{3}\bar P-p_{\xi}+\frac{1}{2}p_{\eta}\right)\,
S^3_{F\,a_3 a'_3}\left(\textstyle \frac{1}{3}\bar P-p_{\eta}\right)\nonumber \\
&& \,\times\, (-\mathrm{i})\,
\int\frac{\mathrm{d}^4 p_\xi'}{(2\pi)^4}\,\frac{\mathrm{d}^4 p_\eta'}{(2\pi)^4}\,
K^{}_{\bar P\, a'_1 a'_2 a'_3;\, a''_1 a''_2 a''_3}
(p_\xi,p_\eta;\, p_\xi',p_\eta')\nonumber \\
&& \quad\quad\quad\quad\quad\quad\quad\quad\quad\quad\quad\quad\quad\times\chi_{\bar P\, a''_1 a''_2 a''_3}(p_\xi',p_\eta'). \label{eqn:BSE}
\end{eqnarray}
Full quark propagators are denoted by $S^i_F$. $K$ stands for the irreducible interaction kernel that contains both irreducible two- and three-particle interactions. The iterative character of the eq.~(\ref{eqn:BSE}) guarantees that an infinite sum of all kinds of reducible diagrams and all irreducible diagrams are considered in the description of the bound state, see fig.~\ref{fig:BSE} for a graphical illustration.
Additionally the Bethe-Salpeter-amplitudes must obey the normalization condition
\begin{eqnarray}
-\mathrm{i}\,\overline\chi_{\bar P}\,\left(P^\mu
\frac{\partial}{\partial P^\mu}(G_{0\,P}^{-1}+\mathrm{i}K_P)\right)\Bigg|_{P=\bar{P}}\chi_{\bar P}=2M^2 \label{eqn:BSA_norm}
\end{eqnarray}
where we have suppressed the integration variables and the multi-indices. Furthermore $M$ is the rest mass of the baryon and $G_{0\,P}$ abbreviates the triple product of the quark propagators $S^i_F$. The normalization condition becomes important in the calculation of matrix elements.

\subsection{Approximations} \label{subsec:approx}
In order to solve eq.~(\ref{eqn:BSE}) we have to make two approximations: Firstly, we use usual quark propagators with constituent quark masses instead of the full propagators $S^i_F$. Secondly, we use instantaneous interactions. It means that in the rest frame of the baryon the irreducible interaction kernel $K_{\bar P}$ does not depend on the relative energies $p^0_\xi$, $p^0_\eta$, $p'\,\!^0_\xi$ and $p'\,\!^0_\eta$. (We do not discuss the complications that arise due to two-particle interactions. This has been elaborated on in great detail in \cite{Loring:2001kv}.) The reduced interaction kernel then reads
\begin{eqnarray}
K^{}_{P}(p^{}_{\xi},p^{}_{\eta},p'_{\xi},p'_{\eta})\Bigg|_{P=(M,\vec
0)}=V^{}(\vec p^{}_{\xi},\vec p^{}_{\eta},\vec
p_{\xi}\,\!\!', \vec p_{\eta}\,\!\!') 
\end{eqnarray}
where again we have suppressed the multi-indices. In our model $V$ contains a phenomenologically adjusted string-like confinement equipped with a suitable spinoral Dirac-structure. Furthermore, $V$ contains 't Hooft's instanton-induced potential. 

With the definition
\begin{eqnarray}
\Phi^{}_M(\vec p^{}_\xi,\vec p^{}_\eta)=\int
\frac{\mathrm{d}p^0_\xi}{2\pi}\frac{\mathrm{d}p^0_\eta}{2\pi}\,\chi^{}_M(p^{}_\xi,p^{}_\eta)
\label{DefSalAmpl}
\end{eqnarray}
for the Salpeter-amplitude it is now possible to eliminate the relative energies in eq.~(\ref{eqn:BSE}). This leads to an eigenvalue problem for the Salpeter-amplitude that can be solved by standard techniques. The eigenvalues obtained are the masses $M$.

With this procedure we are able to produce complete baryon mass spectra up to the highest known energies and spins. These spectra describe very well the experimental situation: Regge-trajectories and hyperfine structures are correctly reproduced. This is done with a very limited number of free parameters, namely seven.

\begin{figure}[t]
\begin{center}
\input{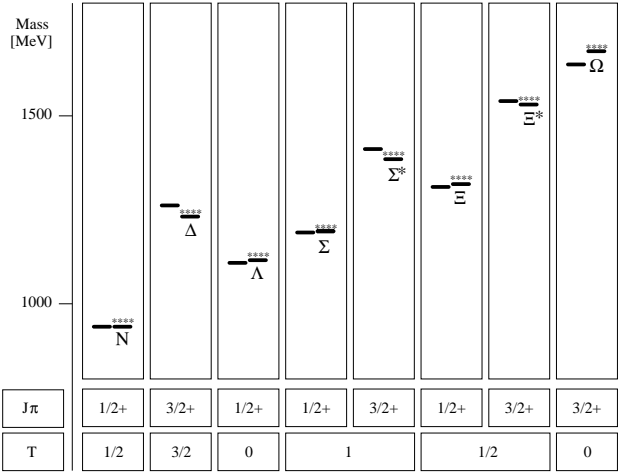}
\caption{The masses of the octet and decuplet ground-state baryons. $J$ denotes total spin, $\pi$ parity and $T$ isospin. On the left side of each column the theoretical values based on the Bethe-Salpeter-model from \cite{Loring:2001kx} and on the right side the experimental masses from \cite{Eidelman:2004wy} are drawn.}
\label{fig:OctDecGround}
\end{center}
\end{figure}

In the present paper we shall concentrate almost exclusively on the octet- and decuplet-baryons displayed in fig.~\ref{fig:OctDecGround}. For more details on the complete light baryon spectra and on the theory of the Bethe-Salpeter-model we refer to \cite{Loring:2001kv,Loring:2001kx,Loring:2001ky}.

\subsection{Current matrix elements}
\begin{figure}[t]
\begin{center}
\input{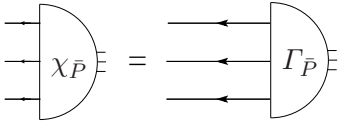}
\caption{Graphical illustration of the definition of the vertex function $\Gamma$ as an amputated Bethe-Salpeter-amplitude $\chi$.}
\label{fig:VertexF}
\end{center}
\end{figure}

The bound state matrix element $\langle\bar P_2|J^\mu(x)|\bar P_1\rangle$ of a current operator
\begin{eqnarray}
J^\mu(x)=: \overline\Psi(x)j^\mu\Psi(x):
\end{eqnarray}
has already been calculated in \cite{Merten:2002nz} using the Mandel\-stam-formalism \cite{Mandelstam:1955sd}. For the initial baryon in its rest frame the matrix element reads
\begin{eqnarray}
&& \hspace{-13pt}\langle\bar P_2|J^\mu(0)|\bar M_1\rangle=
-3\int\frac{\mathrm{d}^4 p_\xi}{(2\pi)^4}\,\frac{\mathrm{d}^4 p_\eta}{(2\pi)^4}
\overline{\Gamma}_{\bar P_2}(\textstyle p_\xi,p_\eta-\frac{2}{3}q)\nonumber \\
&&\hspace{-10pt} S^1_{F}\left(\textstyle \frac{1}{3}\bar M_1+p_{\xi}+\frac{1}{2}p_{\eta}\right)\otimes S^2_{F}\left(\textstyle \frac{1}{3}\bar M_1-p_{\xi}+\frac{1}{2}p_{\eta}\right)\, \nonumber \\
&&\hspace{-10pt}\otimes S^3_{F}\left(\textstyle \frac{1}{3}\bar M_1-p_{\eta}+q\right)\,j^\mu
S^3_{F}\left(\textstyle \frac{1}{3}\bar M_1-p_{\eta}\right)
\Gamma_{\bar M_1}(\textstyle \vec p_\xi,\vec p_\eta). \label{eqn:matrixelement}
\end{eqnarray}
This matrix element contains the vertex function $\Gamma$ defined as an amputated Bethe-Salpeter-amplitude by
\begin{eqnarray}
&& \Gamma_{\bar P}(p_\xi,p_\eta)=
S^{1^{-1}}_{F\,}(\textstyle \frac{1}{3}\bar P+p_{\xi}+\frac{1}{2}p_{\eta})\, \nonumber \\
&& \quad\quad\quad\quad\otimes S^{2^{-1}}_{F}(\textstyle \frac{1}{3}\bar P-p_{\xi}+\frac{1}{2}p_{\eta})\otimes
S^{3^{-1}}_{F}(\textstyle \frac{1}{3}\bar P-p_{\eta}) \nonumber \\
&&\quad\quad\quad\quad\quad\quad\chi_{\bar P}(p_\xi,p_\eta) \label{eqn:VertexF}
\end{eqnarray}
and is depicted in fig.~\ref{fig:VertexF}. Because the Bethe-Salpeter-am\-pli\-tudes are normalized according to eq.~(\ref{eqn:BSA_norm}) the vertex functions are also normalized.

The vertex function depends on the full Bethe-Salpeter-amplitude. As explained in sect.~\ref{subsec:approx} we only calculate the Salpeter-amplitude. Nevertheless we can reconstruct the vertex function by combining eqs.~(\ref{eqn:BSE}) and (\ref{eqn:VertexF}). (We do not discuss the complications during the reconstruction process arising again due to two-particle interactions which is discussed in \cite{Merten:2002nz}.) The reconstruction prescription then reads
\begin{eqnarray}
&& \Gamma_{M}(\vec p_\xi,\vec p_\eta)=
-\mathrm{i}\int\frac{\mathrm{d}^3 p'_\xi}{(2\pi)^3}\,\frac{\mathrm{d}^3 p'_\eta}{(2\pi)^3}\nonumber \\
&&\quad\quad\quad\quad\quad\quad\quad\times V^{}_M(\vec p^{}_{\xi},\vec p^{}_{\eta},\vec
p_{\xi}\,\!\!', \vec p_{\eta}\,\!\!')\Phi^{}_M(\vec p'_\xi,\vec p'_\eta).
\end{eqnarray}
Because we compute the Salpeter-amplitude $\Phi_M$ only in the rest frame of the baryon this determines also the vertex function $\Gamma_M$ in the rest frame. The relativistic covariance of the Bethe-Salpeter-model now allows a correct boosting procedure. It reads
\begin{eqnarray}
\Gamma_{\bar P}(p_\xi,p_\eta)=\bigotimes_{i=1}^3 S_{\Lambda^{\phantom{1}}}\!\Gamma_{\Lambda^{-1}\bar P=(M,\vec{0})}(\vec{\Lambda^{-1}}p_\xi,\vec{\Lambda^{-1}}p_\eta) \label{eqn:boost}
\end{eqnarray}
where $\Lambda$ is a Lorentz-transformation and $S_{\Lambda}$ is the well known representation of the Lorentz-group on Dirac-spi\-nors. By using eq.~(\ref{eqn:boost}) for the description of the outgoing baryon recoil effects are correctly described.

We are thus able to calculate every vertex function needed and finally the matrix element given in eq.~(\ref{eqn:matrixelement}).

\section{Basics of semileptonic decays} \label{sec:BasicsSemilep}
\subsection{Kinematics}
The goal is to calculate semileptonic decays, i.e.~decays of a baryon ground state of a certain mass $M_1$ and four-momentum $\bar P_1$ into another baryon ground state with a smaller mass $M_2$ and four-momentum $\bar P_2$, a lepton with mass $m_{\mathrm{l}}$ and its associated anti-neutrino. Conservation of four-momenta implies
\begin{eqnarray}
\bar P_1=\bar P_2+q \label{eqn:conservation}
\end{eqnarray}
where $q$ is the four-momentum transfer. In a frame where the decaying baryon is at rest eq.~(\ref{eqn:conservation}) becomes
\begin{eqnarray}
\left(\begin{array}{c}
M_1 \\ {\bf 0} 
\end{array} \right)
=
\left(\begin{array}{c}
\sqrt{M_2^2+ {\bf P}^2} \\ {\bf P} 
\end{array} \right)
+
\left(\begin{array}{c}
M_1-\sqrt{M_2^2+ {\bf P}^2} \\ -{\bf P} 
\end{array} \right)
\end{eqnarray}
with ${\bf P}$ denoting the three-momentum of the final baryon. For minimal ${\bf P}^2$
we have maximal $q^2$ and vice versa. For 
\begin{eqnarray}
{\bf P}^2=0
\end{eqnarray}
we have
\begin{eqnarray}
q^2_\mathrm{max}=(M_1-M_2)^2 \label{eqn:sem_maxq}
\end{eqnarray}
and ${\bf P}^2$ is maximal for
\begin{eqnarray}
M_1-\sqrt{M_2^2+ {\bf P}^2}=\sqrt{m_{\mathrm{l}}^2+{\bf P}^2}.
\end{eqnarray}
Then we have
\begin{eqnarray}
q^2_\mathrm{min}=m^2_{\mathrm{l}}.
\end{eqnarray}
Every matrix element depending on $q^2$ is thus a function with arguments between $q^2_\mathrm{min}$ and $q^2_\mathrm{max}$.

\subsection{Helicity amplitudes}
The semileptonic decay occurs via virtual W-boson exchange coupling to a quark. To implement this coupling we use the standard weak vector and axial vector current $J_{\mu}^{V+A}$.
As in ref.~\cite{Korner:1991ph} we introduce four $q^2$-dependent and invariant vector and axial form factors $F^{V/A}_{1/2}$ defined by
\begin{eqnarray}
\langle\bar P_2|J_\mu^{V+A}|\bar P_1\rangle =&&\bar
u(\bar P_2)[\gamma_\mu(F_1^V+F_1^A\gamma_5)\nonumber
\\&&+\mathrm{i}\sigma_{\mu\nu}q^\nu(F_2^V+F_2^A\gamma_5)]u(\bar P_1) \label{eqn:formf}
\end{eqnarray}
where $u(\bar P_i)$ are Dirac-spinors normalized as $u^{\dagger}u=2E$ and
\begin{eqnarray}
\sigma_{\mu\nu}=\frac{\mathrm{i}}{2}(\gamma_\mu\gamma_\nu-\gamma_\nu\gamma_\mu).
\end{eqnarray}
For the time being we have neglected invariants multiplying $q_\mu$
which is justified in the zero-lepton-mass limit. This is indeed a very good approximation when we consider electrons but will be corrected below when we deal with muonic decays.

The corresponding helicity amplitudes $H^{V/A}_{\lambda_2\,\lambda_W}$ are as in \cite{Korner:1991ph} defined through the form factors by
\begin{eqnarray}
\sqrt{q^2}{H^{V/A}_{1/2\,0}}=\sqrt{Q_{\mp}}[(M_1\pm
M_2){F_1^{V/A}}\mp q^2{F_2^{V/A}}] \label{eqn:hellong}
\end{eqnarray}
and
\begin{eqnarray}
{H^{V/A}_{1/2\,1}}=\sqrt{2Q_{\mp}}[-{F_1^{V/A}}\pm(M_1\pm
M_2){F_2^{V/A}}] \label{eqn:heltrans}
\end{eqnarray}
where $\lambda_2$ and $\lambda_W$ are the helicities of the final
baryon and the W-boson respectively. Moreover the abbreviation
\begin{eqnarray}
Q_\pm=(M_1\pm M_2)^2-q^2
\end{eqnarray}
is used. The remaining helicity amplitudes with negative $\lambda_2$ are obtained through the parity relations
\begin{eqnarray}
{H^{V/A}_{-\lambda_2\,-\lambda_W}}=\pm {H^{V/A}_{\lambda_2\,\lambda_W}}. \label{eqn:hel_parity}
\end{eqnarray}

We can now eliminate the form factors from eqs.~(\ref{eqn:hellong})
and (\ref{eqn:heltrans}) with the help of
eq.~(\ref{eqn:formf}). With the notation
\begin{eqnarray}
H_{\lambda_2\,\lambda_W}=H^V_{\lambda_2\,\lambda_W}+H^A_{\lambda_2\,\lambda_W} \label{eqn:hel_complete}
\end{eqnarray}
for the complete helicity amplitude $H_{\lambda_2\,\lambda_W}$ this leads to
\begin{eqnarray}
\sqrt{q^2}{H_{\lambda_2\,0}}=&&\langle\bar P_2|J_0^{V+A}|\bar
P_1\rangle\cdot(-q^3)\nonumber\\&&-\langle\bar P_2|J_3^{V+A}|\bar
P_1\rangle\cdot q^0 \label{eqn:translong}
\end{eqnarray}
and
\begin{eqnarray}
\pm{H_{\lambda_2\,\pm 1}}=\langle\bar P_2|J^{V+A}_{\pm}|\bar P_1\rangle \label{eqn:transtrans}
\end{eqnarray}
with the definitions
\begin{eqnarray}
J^{V+A}_{\pm}=\frac{1}{\sqrt{2}}(J_1^{V+A}\pm\mathrm{i}J_2^{V+A}).
\end{eqnarray}
The $z$-axis is chosen along the momentum transfer.

\begin{figure}[t]
\begin{center}
\input{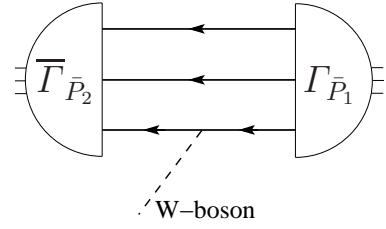}
\caption{Graphical illustration of the calculated weak current induced baryonic transition according to eq.~(\ref{eqn:matrixelement}). Within the Mandelstam-formalism in lowest order all strong interactions have been absorbed into the vertex functions $\Gamma_{\bar P_i}$ leaving only propagators and a weak coupling.}
\label{fig:sem_trans}
\end{center}
\end{figure}

The right-hand sides of eqs.~(\ref{eqn:translong}) and (\ref{eqn:transtrans}) contain the baryonic transition amplitudes to be calculated within our Bethe-Salpeter-formalism using eq.~(\ref{eqn:matrixelement}). This is diagrammatically illustrated in fig.~\ref{fig:sem_trans}. 

\subsection{Observables}
Finally we give the expressions for the quantities depending on the helicity amplitudes for which experimental data are available: The first quantity is the axial-vector-to-vec\-tor form factor ratio $g_A/g_V$. Eliminating the induced tensor contributions $F_2^{V/A}$ from eqs.~(\ref{eqn:hellong}) and (\ref{eqn:heltrans}) leads to
\begin{eqnarray}
\frac{F_1^A}{F_1^V}=\frac{[\sqrt{2}(M_1-M_2)H_{1/2\,0}^A+\sqrt{q^2}H_{1/2\,1}^A]\sqrt{Q_+}}{[\sqrt{2}(M_1+M_2)H_{1/2\,0}^V+\sqrt{q^2}H_{1/2\,1}^V]\sqrt{Q_-}}. \label{eqn:gAgV}
\end{eqnarray}
For $q^2\to 0$ eq.~(\ref{eqn:gAgV}) becomes (in the notation of \cite{Eidelman:2004wy})
\begin{eqnarray}
\frac{g_A}{g_V}=\frac{g_1(0)}{f_1(0)}=\frac{F_1^A(0)}{F_1^V(0)}=\frac{H_{1/2\,0}^A(0)}{H_{1/2\,0}^V(0)}.
\end{eqnarray}
We notice that this ratio does not depend directly on the baryon masses nor on the four-momentum transfer anymore but on the helicity amplitudes alone.

The next quantity is the weak-magnetism form factor ratio $f_2(0)/f_1(0)$. Using the vector contributions $F_{1/2}^V$ of eqs.~(\ref{eqn:hellong}) and (\ref{eqn:heltrans}) one finds after some rearrangement 
\begin{eqnarray}
\frac{f_2(q^2)}{f_1(q^2)}=\frac{M_1\sqrt{2q^2}H_{1/2\,0}^V+M_1(M_1+M_2)H_{1/2\,1}^V}{q^2H_{1/2\,1}^V+\sqrt{2q^2}(M_1+M_2)H_{1/2\,0}^V} \label{eqn:f2f1}
\end{eqnarray}
from which the ratio $f_2(0)/f_1(0)$ can then be obtained.

For the two form factor ratios we need the vector and the axial parts of the helicity amplitudes separately. They can be obtained by combining eqs.~(\ref{eqn:hel_parity}) and (\ref{eqn:hel_complete}).

The last quantity that we consider in this paper is the decay rate $\Gamma$ that can be obtained using
standard techniques. After integrating out the angular dependence of the decay distribution the theoretical $q^2$-dependent differential decay rate reads
\begin{eqnarray}
\frac{\mathrm{d}{\Gamma}}{\mathrm{d}q^2}=\frac{G^2_{\mathrm{F}}}{(2\pi)^3}|V_{q_1q_2}|^2\frac{q^2p}{24M_1^2}\sum_{\lambda_2,
 \lambda_W}|{H_{\lambda_2\,\lambda_W}}|^2 \label{eqn:semDiffRate}
\end{eqnarray}
where $G_\mathrm{F}$ is the Fermi-coupling constant, $V_{q_1q_2}$ is the Ca\-bibbo-Kobayashi-Maskawa-matrix element for the quark transition $q_1\to q_2$ and
\begin{eqnarray}
p=\frac{\sqrt{Q_+Q_-}}{2M_1}. \label{eqn:sem_kinfac}
\end{eqnarray}

Because we are mainly interested in the quality of our transition matrix elements given in eqs.~(\ref{eqn:translong}) and (\ref{eqn:transtrans}), we insert the experimental and not our theoretical values for the baryon masses $M_1$ and $M_2$ into eqs.~(\ref{eqn:f2f1}) and (\ref{eqn:semDiffRate}). The helicity amplitudes remain untouched by this fine-tuning procedure.

\subsection{Non-zero-lepton-mass}
Experimental data on decay rates for muonic semileptonic baryon decays are sparse but available. In the non-zero-lepton-mass limit the parameterization in eq.~(\ref{eqn:formf}) has to be extended like in \cite{Pietschmann:1983wi} to the form
\begin{eqnarray}
\langle\bar P_2|J_\mu^{V+A}|\bar P_1\rangle =&&\bar
u(\bar P_2)[\gamma_\mu(F_1^V+F_1^A\gamma_5)\nonumber
\\&&+\mathrm{i}\sigma_{\mu\nu}q^\nu(F_2^V+F_2^A\gamma_5)\nonumber \\&&+q_\mu(F_3^V+F_3^A\gamma_5)]u(\bar P_1)
\end{eqnarray}
where the additional form factors $F_3^{V/A}$ appear.

The term for the differential decay rate on the right hand side of eq.~(\ref{eqn:semDiffRate}) has also to be modified. From \cite{Pietschmann:1983wi} we adopt a lengthy expression for the decay rate where in its derivation the lepton mass has been kept finite. This expression depends on the six form factors $F_{1,2,3}^{V/A}$ and on the participating masses. By introducing the additional helicity amplitude 
\begin{eqnarray}
\sqrt{q^2}{H'_{\lambda_2\,0}}=&&\langle\bar P_2|J_0^{V+A}|\bar
P_1\rangle\cdot(-q^0)\nonumber\\&&-\langle\bar P_2|J_3^{V+A}|\bar
P_1\rangle\cdot q^3 
\end{eqnarray}
the differential decay rate in \cite{Pietschmann:1983wi} can be written as
\begin{eqnarray}
\frac{\mathrm{d}{\Gamma}}{\mathrm{d}q^2}=&&\frac{G^2_{\mathrm{F}}}{(2\pi)^3}|V_{q_1q_2}|^2\frac{q^2p}{24M_1^2}\frac{(q^2-m_{\mathrm{l}}^2)^2}{(q^2)^2}\nonumber\\
&&\times\Big[(1+\frac{m_\mathrm{l}^2}{2q^2})\sum_{\lambda_2,
\lambda_W}|{H_{\lambda_2\,\lambda_W}}|^2\nonumber \\
&&+\frac{3m_\mathrm{l}^2}{2q^2}\sum_{\lambda_2, \lambda_W}|{H'_{\lambda_2\,0}}|^2\Big], \label{eqn:semDiffRate_muon}
\end{eqnarray}
see also ref.~\cite{Kadeer:2005aq} in which eq.~(\ref{eqn:semDiffRate_muon}) is explicitly derived.

We now have all the tools to compute electronic and muonic semileptonic decays of baryons within the Bethe-Salpeter-model.

\section{Results and discussion} \label{sec:results}

\subsection{General remarks}
In the calculation of transition amplitudes in the Mandel\-stam-formalism in lowest order we do not need to introduce additional free parameters. The very limited number of parameters that enter into the Bethe-Salpeter-model have been fixed by the baryon mass spectra alone \cite{Loring:2001kv,Loring:2001kx,Loring:2001ky}. We want to stress that every calculation that is presented from this point on is therefore a genuine prediction. 

We put special emphasis on the decays where the incoming and the outgoing ground state baryons are both spin-$\frac{1}{2}$ particles, these are the baryons $N$, $\Lambda$, $\Sigma$ and $\Xi$. This is the most important case since nearly all experimental data are within this sector. We have exact SU(2) isospin symmetry.  After taking into account the kinematical restrictions and the fact that strangeness cannot change by more than one unit only five essentially different types of transitions are left, namely $\Lambda\to N$, $\Sigma\to N$, $\Sigma\to\Lambda$, $\Xi\to\Lambda$ and $\Xi\to\Sigma$. In addition we calculate the transition $\Lambda_c\to\Lambda$. The $\Lambda_c$ has been calculated using also a phenomenological extension of 't Hooft's force to charmed quarks as has been done for mesons before \cite{Merten:2001er}. This is done to keep the model simple and universal. In this way the experimental mass of the $\Lambda_c$ is very well reproduced. It turned out that this extension of 't Hooft's interaction is indeed sufficient to describe charmed baryons. We thus even do not use the one-gluon exchange when we deal with heavy baryons. We take a closer look at this matter in ref.~\cite{Migura:2006ep}.

For decays with decuplet-baryons as incoming particles additional helicity amplitudes $H_{1/2\,-1}$ and $H_{-1/2\,1}$  enter into eqs.~(\ref{eqn:semDiffRate}) and (\ref{eqn:semDiffRate_muon}). The transition $\Delta\to\Lambda$ is not possible because isospin cannot change by more than one unit.

\subsection{Form factor ratios}
With eqs.~(\ref{eqn:gAgV}) and (\ref{eqn:f2f1}) we first calculate the form factor ratios $F_1^A(q^2)/F_1^V(q^2)$ and $f_2(q^2)/f_1(q^2)$. Experimental values are only available for $q^2=0$ and are compared in the last four columns in table \ref{tab:results} to our predictions.
\begin{table*}[]
\begin{center}
\begin{tabular}{|r@{ $\to$ }l||r@{.}l|r@{ $\pm$ }l||r@{.}l|r@{ $\pm$ }l||r@{.}l|r@{ $\pm$ }l||r@{.}l|r@{ $\pm$ }l|}
\hline
\multicolumn{2}{|c||}{}&\multicolumn{4}{|c||}{$\Gamma, l=e$}&\multicolumn{4}{|c||}{$\Gamma, l=\mu$}&\multicolumn{4}{|c||}{$g_A/g_V$}&\multicolumn{4}{|c|}{$f_2(0)/f_1(0)$} \\
\hline
\multicolumn{2}{|c||}{Decay}&\multicolumn{2}{|c|}{BSM}&\multicolumn{2}{|c||}{EXP}&\multicolumn{2}{|c|}{BSM}&\multicolumn{2}{|c||}{EXP}&\multicolumn{2}{|c|}{BSM}&\multicolumn{2}{|c||}{EXP}&\multicolumn{2}{|c|}{BSM}&\multicolumn{2}{|c|}{EXP}\\
\hline
\hline
$\Lambda^0$&$pl^{-}\bar\nu_l$			&3&15	&3.16&0.06	&0&51	&0.60&0.13	&$-0$&82	&$-0.718$&0.015&$-0$&78&\multicolumn{2}{|c|}{}\\
$\Sigma^-$&$nl^{-}\bar\nu_l$			&4&24	&6.88&0.24	&1&79	&3.04&0.27	&0&25	& 0.340&0.017&0&91&0.97&0.14\\
$\Sigma^-$&$\Lambda^0l^{-}\bar\nu_l$		&0&35	&0.39&0.02	&\multicolumn{2}{|c|}{-}&\multicolumn{2}{|c||}{-}&0&005	& 0.01& 0.10&0&003&\multicolumn{2}{|c|}{}\\
$\Sigma^+$&$\Lambda^0l^{+}\nu_l$		&0&21	&0.25&0.06	&\multicolumn{2}{|c|}{-}&\multicolumn{2}{|c||}{-}&0&005	&\multicolumn{2}{|c||}{}&0&003&\multicolumn{2}{|c|}{}\\
$\Xi^-$&$\Lambda^0l^{-}\bar\nu_l$		&2&56	&3.44&0.19	&0&73	&\multicolumn{2}{|c||}{$2.14^{+2.14}_{-1.34}$}&$-$0&27	&$-0.25$& 0.05&0&026&\multicolumn{2}{|c|}{}\\
$\Xi^0$&$\Sigma^+l^{-}\bar\nu_l$		&0&90	&0.93&0.14	&0&004	&\multicolumn{2}{|c||}{$0.016^{+0.007}_{-0.006}$}&$-1$&38	&\multicolumn{2}{|c||}{$-1.32_{-0.22}^{+0.18}$}&$-1$&93&$-2.0$&1.3\\
$\Xi^-$&$\Sigma^0l^{-}\bar\nu_l$		&0&50	&0.51&0.10	&0&003	&\multicolumn{2}{|c||}{$<4.8$}&$-1$&38	&\multicolumn{2}{|c||}{}&$-1$&94&\multicolumn{2}{|c|}{}\\
$\Lambda_c^+$&$\Lambda^0l^{+}\nu_l$		&1&57	&1.05&0.30	&1&46	&1.00&0.35	&$-0$&95	&\multicolumn{2}{|c||}{}&$-0$&53&$-0.31$&0.06\\
$\Omega^-$&$\Xi^0l^{-}\bar\nu_l$		&\multicolumn{2}{|c|}{92}	&68&34	&\multicolumn{2}{|c|}{59}	&\multicolumn{2}{|c||}{}	&\multicolumn{2}{|c|}{}	&\multicolumn{2}{|c||}{}&\multicolumn{2}{|c|}{}&\multicolumn{2}{|c|}{}\\
\hline
\end{tabular}
\end{center}
\caption{Theoretical (BSM standing for Bethe-Salpeter-Model) and experimental (EXP) values for semileptonic baryon decays. The participating lepton $l$ can either be an electron $e$ or a muon $\mu$. The decay rates $\Gamma$ are given in $10^6 s^{-1}$, with the exception of the decays $\Lambda_c^+\to\Lambda^0l^{+}\nu_l$ where the numbers are given in $10^{11} s^{-1}$. The experimental values for the decay rates $\Gamma$ are estimates by the authors of \cite{Eidelman:2004wy}, with the exception of the decay $\Xi^0\to\Sigma^+\mu^-\bar\nu_\mu$ which is from \cite{Alavi-Harati:2005ev}. A dash indicates that the decay is kinematically not possible. Also presented are the theoretical predictions and experimental measurements of the axial-vector-to-vector coupling ratio $g_A/g_V$ and weak magnetism ratio $f_2(0)/f_1(0)$. Note that we use the sign convention of \cite{Eidelman:2004wy} and note that for $\Sigma\to\Lambda$ the reciprocals are given. The experimental data for the decay $\Xi^0\to\Sigma$ is from \cite{Alavi-Harati:2001xk} and for $\Lambda_c^+\to\Lambda$ from \cite{Hinson:2004pj}, the other experimental values are again from \cite{Eidelman:2004wy}.} \label{tab:results}
\end{table*}

The experimental values for the form factor ratios main\-ly date from the eighties. One exception is the transition $\Xi^0\to\Sigma^+$ observed and measured in 2001 by the KTeV Collaboration at Fermilab \cite{Alavi-Harati:2001xk}. The other exception is the very recent measurement of the transition $\Lambda^+_c\to\Lambda$ in 2005 by the CLEO Collaboration at the Cornell Electron Storage Ring \cite{Hinson:2004pj}. All experimental values correspond to electronic semileptonic decays.

Comparing our predictions with the experimental form factor ratios we find a good agreement. Not a single value is totally off the mark. Five out of eight theoretical numbers lie within the error bars. The other three predictions have the right sign and the correct order of magnitude but are somewhat too small: for $g_A/g_V$ the transitions $\Lambda^0\to p$ and $\Sigma^-\to n$ and for $f_2(0)/f_1(0)$ the transition $\Lambda^+_c\to\Lambda^0$.

In addition we present in table \ref{tab:results} eight pure predictions without an experimental counterpart. In particular, it will be interesting if $g_A/g_V$ for the decay $\Lambda_c^+\to\Lambda^0e^{+}\nu_e$ is measured in the near future.

\subsection{Decay rates}
\begin{figure}[]
\resizebox{0.48\textwidth}{!}{ 
\includegraphics[viewport = 143 494 506 719]{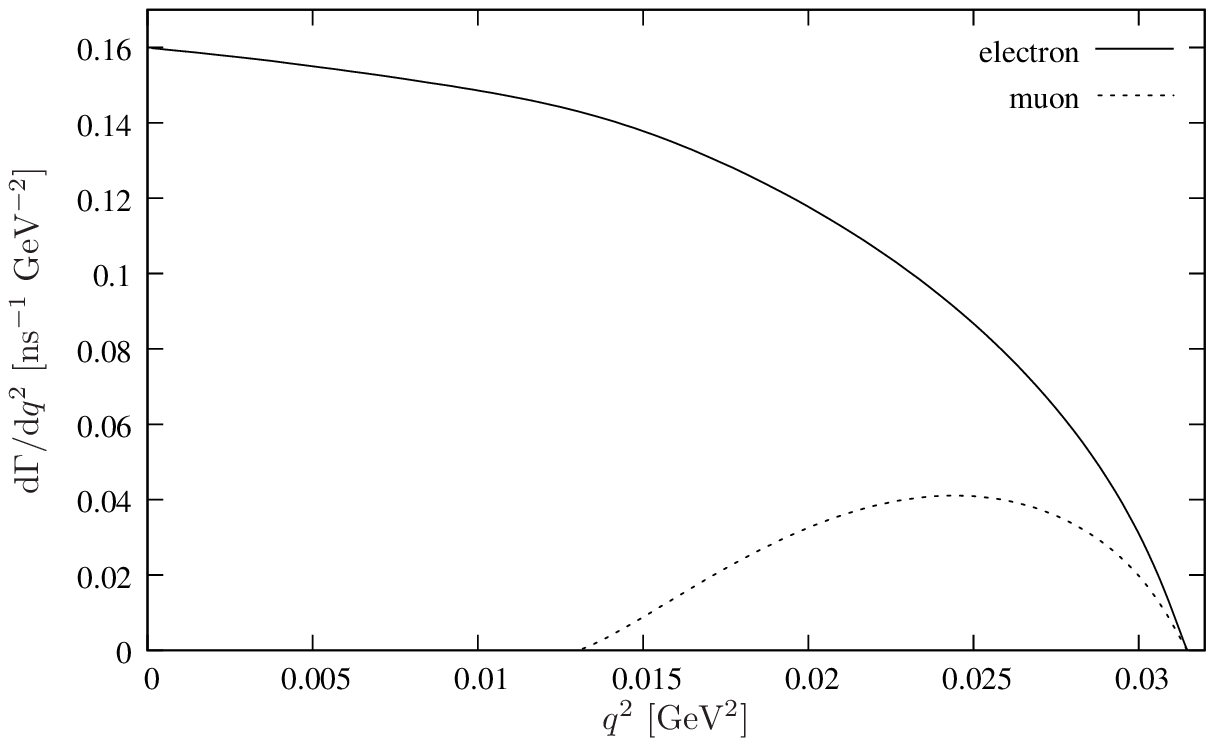}}
\caption{Calculated differential decay rate for $\Lambda^0\to p l^{-} \bar\nu_{l}$ for $l=e$ (solid) and $l=\mu$ (dotted).}
\label{fig:LamProt_dwq}
\end{figure}

\begin{figure}[]
\resizebox{0.48\textwidth}{!}{ 
\includegraphics[viewport = 143 494 506 719]{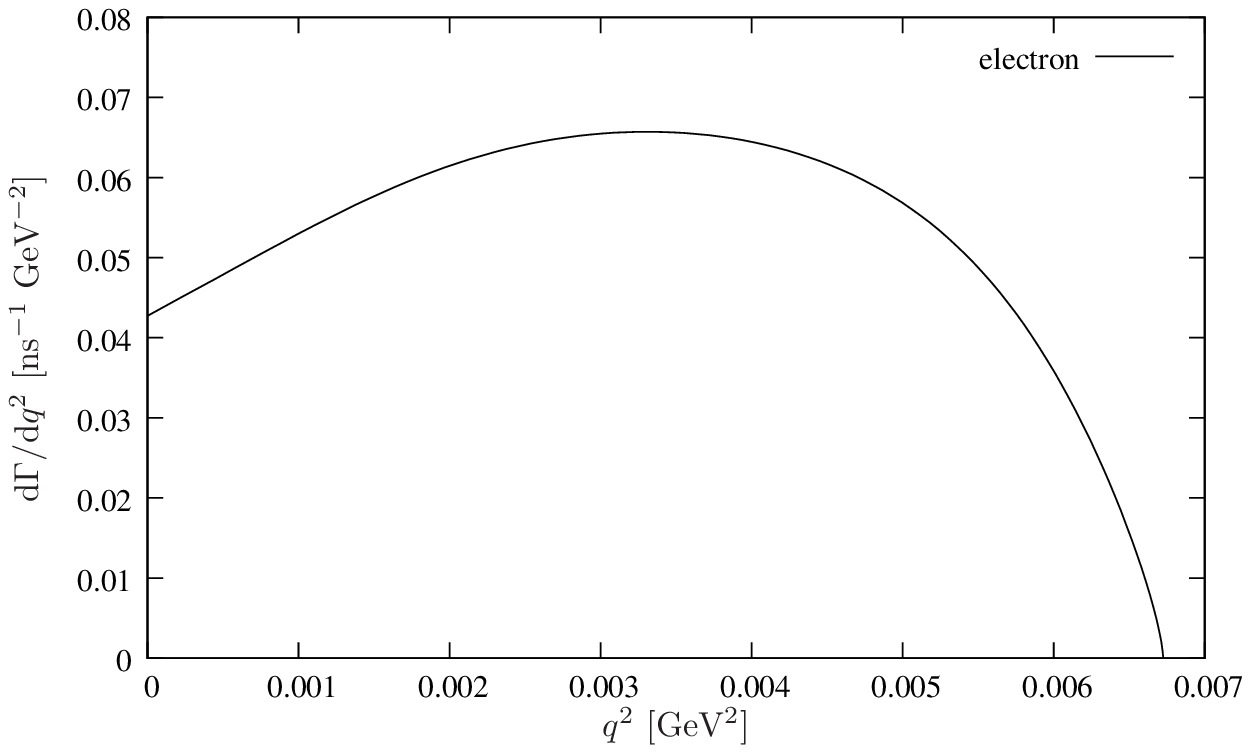}}
\caption{Calculated differential decay rate for $\Sigma^-\to \Lambda^0 e^{-} \bar\nu_{e}$ for $l=e$ (solid) and $l=\mu$ (dotted).}
\label{fig:SigLam_dwq}
\end{figure}

\begin{figure}[]
\resizebox{0.48\textwidth}{!}{ 
\includegraphics[viewport = 143 494 506 719]{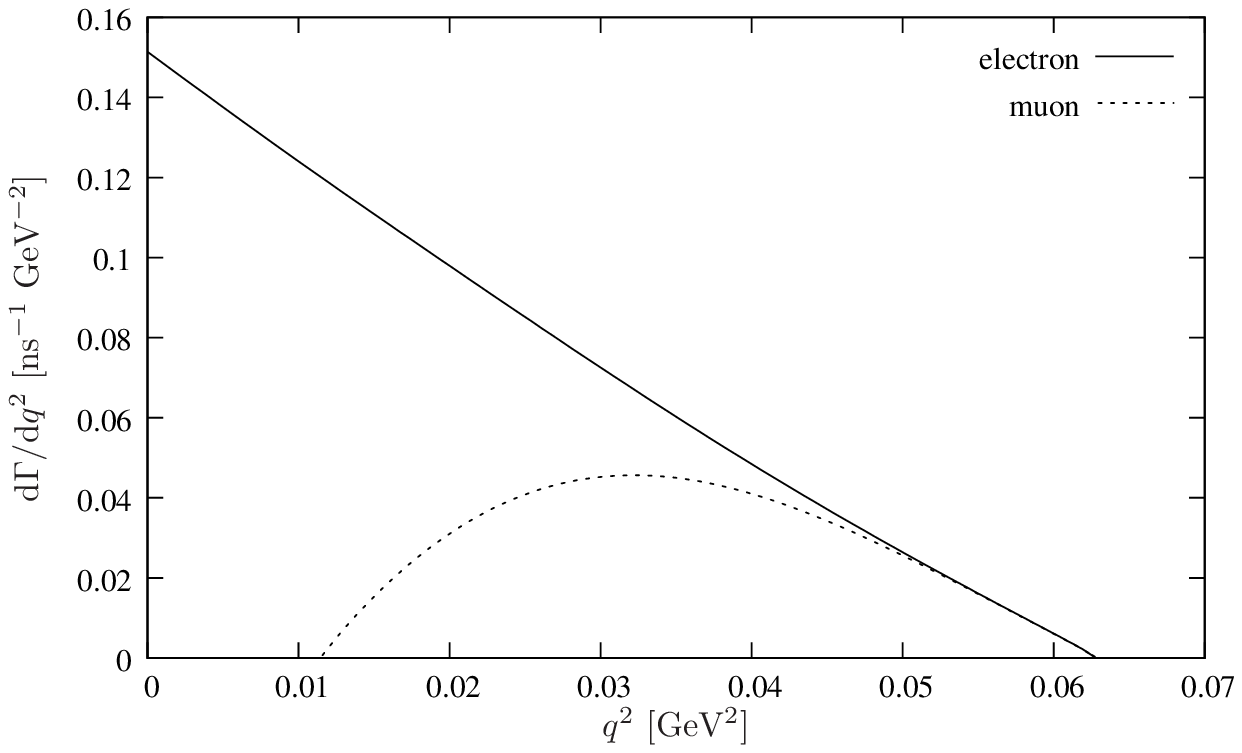}}
\caption{Calculated differential decay rate for $\Sigma^-\to n l^{-} \bar\nu_{l}$ for $l=e$ (solid) and $l=\mu$ (dotted).}
\label{fig:SigNeu_dwq}
\end{figure}

\begin{figure}[]
\resizebox{0.48\textwidth}{!}{ 
\includegraphics[viewport = 143 494 506 719]{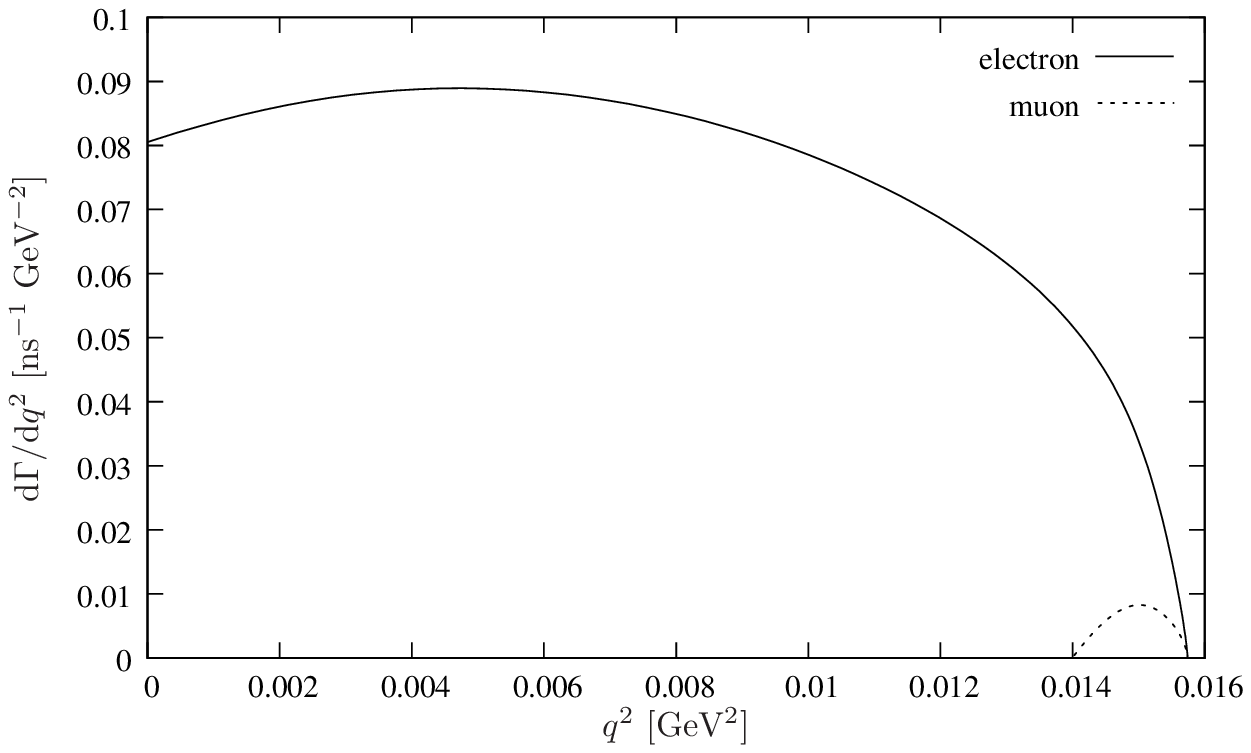}}
\caption{Calculated differential decay rate for $\Xi^0\to \Sigma^+ l^{-} \bar\nu_{l}$ for $l=e$ (solid) and $l=\mu$ (dotted). The values for the muonic curve have been multiplied by 50.}
\label{fig:XiSig_dwq}
\end{figure}

\begin{figure}[]
\resizebox{0.48\textwidth}{!}{ 
\includegraphics[viewport = 143 494 506 719]{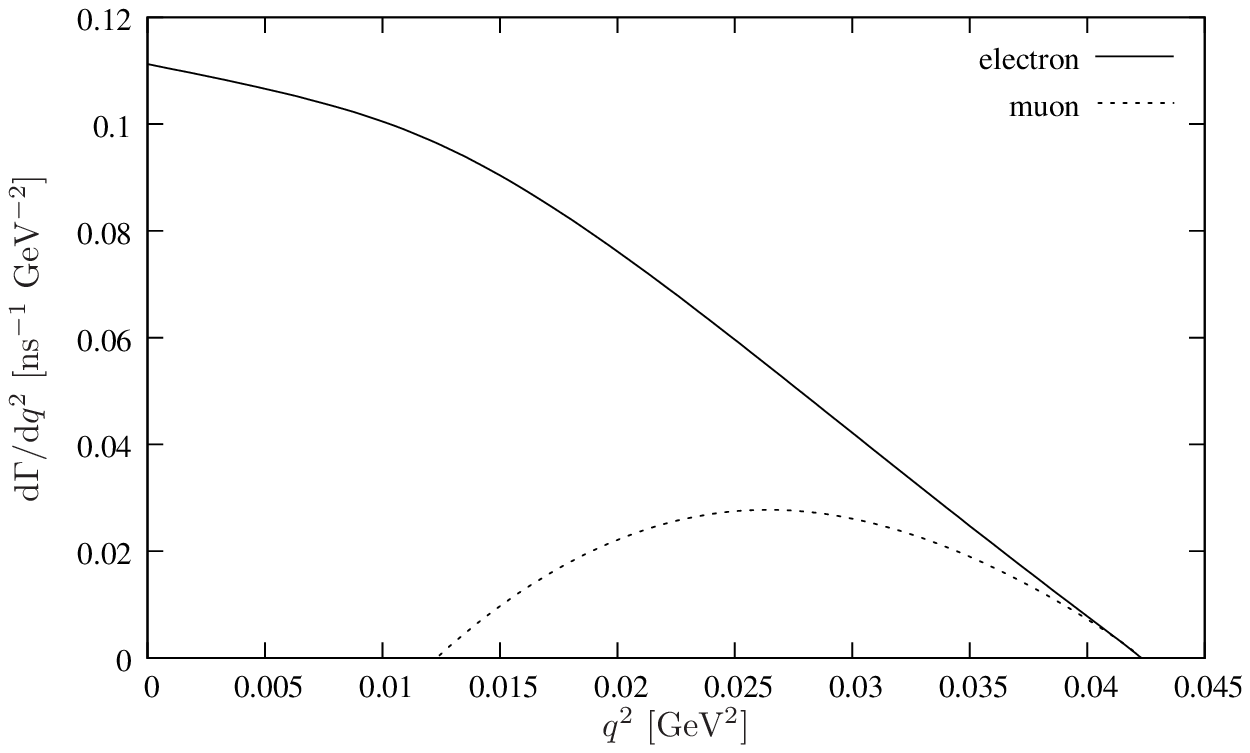}}
\caption{Calculated differential decay rate for $\Xi^-\to \Lambda^0 l^{-} \bar\nu_{l}$ for $l=e$ (solid) and $l=\mu$ (dotted).}
\label{fig:XiLam_dwq}
\end{figure}

\begin{figure}[]
\resizebox{0.48\textwidth}{!}{ 
\includegraphics[viewport = 143 494 506 719]{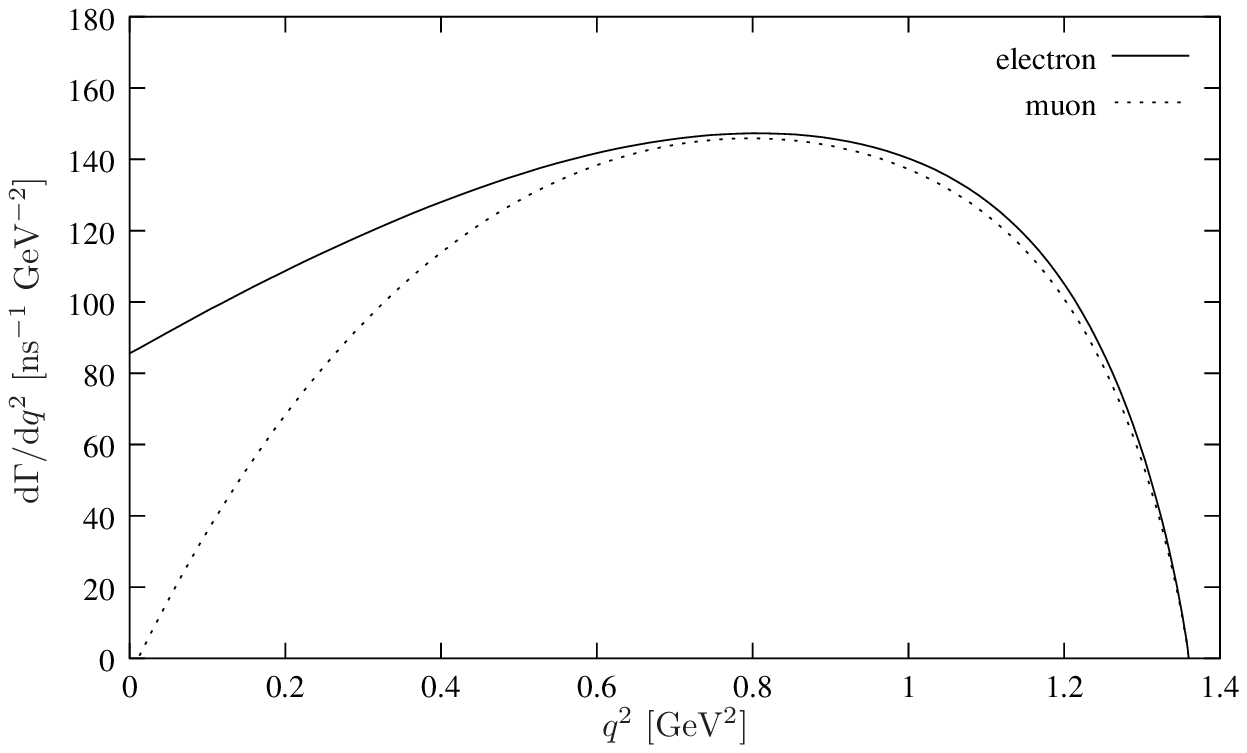}}
\caption{Calculated differential decay rate for $\Lambda^+_{C} \to \Lambda^0 l^{+} \nu_{l}$ for $l=e$ (solid) and $l=\mu$ (dotted).}
\label{fig:LamCLam_dwq}
\end{figure}

To obtain the decay rates we first have to calculate the differential decay rates. Equation (\ref{eqn:semDiffRate_muon}) allows us to compute the differential decay rates for electronic and muonic semileptonic decays. The results are presented in figs.~\ref{fig:LamProt_dwq}-\ref{fig:LamCLam_dwq}.

These differential decay rates are pure theoretical predictions without any direct experimental confirmation so far. To our knowledge there exist none whereas there exist some data for semileptonic decays of mesons \cite{Barish:1994mu,Athanas:1997eg}.

The global behavior of the curves is very much influenced by the phase factor proportional to $q^2p$. In the case of electrons one would maybe naively expect that the differential decay rate vanishes for $q^2=0$. That is not true due to the limit behavior of the helicity amplitudes. But the differential decay rate indeed becomes zero at $q^2_{\mathrm{min}}=m^2_\mu$ for muons.

Integrating these curves one obtains the decay rates which are compared to known experimental numbers in table \ref{tab:results}.

First measurements of semileptonic decay rates go back to the early sixties. But the decisive measurements of electronic semileptonic decays with lots of events were done in the eighties using the charged hyperon beam at the CERN Super Proton Synchrotron \cite{Bourquin:1981ba}. The most recent experiments are the observations of $\Xi^0\to\Sigma^+ l^-\bar\nu_l$ by the KTeV Collaboration in 1999 and 2005 \cite{Affolder:1999pe,Alavi-Harati:2005ev}. The muonic decays have naturally been seen in very few events. An extreme example is the transition $\Xi^-\to\Lambda^0$ where merely a single event has been observed only. 

If we compare our predictions and the experimental decay rates we find again good overall agreement. Not one single value is totally off the mark. Eleven out of fifteen theoretical numbers lie virtually within the error bars. Only four theoretical values are clearly too small, namely the rates for both decays $\Sigma^-\to nl^-\bar\nu_l$, for the decay $\Xi^-\to \Lambda^0e^-\bar\nu_e$ and for the decay $\Xi^0\to\Sigma^+ \mu^-\bar\nu_\mu$. Nevertheless they have the correct order of magnitude. The only experimental values that are slightly overestimated are the decay rates for the $\Lambda_c^+$.

\begin{center}
\begin{table}[]
\begin{center}
\begin{tabular}{|r@{ $\to$ }l||r@{.}l|r@{.}l|}
\hline
\multicolumn{2}{|c||}{Decay}&\multicolumn{2}{|c|}{$l=e$}&\multicolumn{2}{|c|}{$l=\mu$}\\
\hline
\hline
$\Delta^{++}$&$p l^{+}\nu_l$	&\multicolumn{2}{|c|}{278}&\multicolumn{2}{|c|}{144}	\\
$\Delta^{++}$&$\Sigma^+ l^{+}\nu_l$	&0&002&\multicolumn{2}{|c|}{-}	\\
$\Sigma^{\ast\,0}$&$p l^{-}\bar\nu_l$	&\multicolumn{2}{|c|}{34}&\multicolumn{2}{|c|}{26}	\\
$\Sigma^{\ast\,0}$&$\Sigma^+ l^{-}\bar\nu_l$	&3&93	&0&80	\\
$\Sigma^{\ast\,0}$&$\Xi^- l^{+}\nu_l$	&0&0008	&\multicolumn{2}{|c|}{-}	\\
$\Xi^{\ast\,-}$&$\Lambda^0 l^{-}\bar\nu_l$	&19&1	&\multicolumn{2}{|c|}{-}	\\
\hline
\end{tabular}
\end{center}
\caption{Predictions of decay rates $\Gamma$ (in $10^6 s^{-1}$) of semileptonic baryon decays based on the Bethe-Salpeter-Model. The participating lepton $l$ can either be an electron $e$ or a muon $\mu$. A dash indicates that the decay is kinematically not possible.} \label{tab:pred_decu}
\end{table}
\end{center}

In addition we present more predictions for semi\-lep\-tonic decays in table \ref{tab:pred_decu}. We note the large decay rate for the transitions $\Delta^{++}\to p$ which is due to the fact that here the Cabibbo-Kobayashi-Maskawa-matrix element $V_{ud}$ is involved whose square is approximately 20 times larger than the square of $V_{us}$. The latter enters in almost all the other decays considered in this paper.

\section{Conclusion} \label{sec:conclusion}
In the framework of a relativistically covariant constituent quark model we computed the semileptonic decays of light-flavored baryons and $\Lambda_c$. We used the Bethe-Salpeter-equation in instantaneous approximation. As interactions we inserted a linearly rising confinement and 't Hooft's instanton-induced interaction. This Bethe-Salpeter-model has already given an accurate description of a variety of experimental data based on a minimal number of free parameters which were completely fixed by the baryon spectra.

In this paper we have added to this results for weak decays, extending the unified picture without introducing additional free parameters. We have computed in succession helicity amplitudes, axial-vector-to-vector coupling ratios, weak magnetism ratios, differential decay rates and finally total decay rates of electronic and muonic semileptonic baryon decays. We have compared our theoretical predictions with every known experimental number and in general found a good agreement. No real contradictions between the Bethe-Salpeter-model and the experiments have been found. 

We have given many predictions of observables for which currently no experimental results exist. There clear\-ly exists an experimental deficit regarding differential decay rates of semileptonic baryon decays. We particularly hope for a confirmation of the axial-vector-to-vector coupling ratio $g_A/g_V=-0.95$ which we predict for the decay $\Lambda_c^+\to\Lambda^0e^+\bar\nu_e$.

 At the energies considered in this paper we have not found overwhelming indication that our ansatz fails, thus leading to the conclusion that the concept of constituent quarks remains successful also for a variety of weak observables. Further tests in the same spirit of a unified approach such as extensive calculations of strong two-body baryon decays are in preparation.

\section*{Acknowledgments}
Financial support from the Deutsche Forschungsgemeinschaft by the
SFB/Transregio 16 ``Subnuclear Structure of Matter'' and from the European
Community-Research Infrastructure Activity under the FP6 "Structuring the
European Research Area" programme (HadronPhysics, contract number
RII3-CT-2004-506078) is gratefully acknowledged.

\end{document}